# Using resource graphs to represent conceptual change

Michael C. Wittmann
Department of Physics, University of Maine
Orono ME 04469-5709
wittmann@umit.maine.edu

## Abstract

We introduce resource graphs, a representation of linked ideas used when reasoning about specific contexts in physics. Our model is consistent with previous descriptions of resources and coordination classes. It can represent mesoscopic scales that are neither knowledge-in-pieces or large-scale concepts. We use resource graphs to describe several forms of conceptual change: incremental, cascade, wholesale, and dual construction. For each, we give evidence from the physics education research literature to show examples of each form of conceptual change. Where possible, we compare our representation to models used by other researchers. Building on our representation, we introduce a new form of conceptual change, differentiation, and suggest several experimental studies that would help understand the differences between reform-based curricula.

## 1. INTRODUCTION

In this paper, we present a simple representation of student understanding of physics that allows us to represent many different types of learning which have been observed and described in the physics education research literature. The representation is a resource graph, in which we represent the small-scale ideas[1-10] which students use in a particular setting as linked to each other and activated by some observation. Our research graph representation is consistent with a description of coordination classes[11] but uses the idea of *causal net* and *readout strategy* at a much smaller, mesoscopic scale. Our representation builds a bridge between a microscopic "resources" level and the macroscopic "concepts" level of describing student reasoning.

We have three goals in creating a useful representation of student thinking. First, we believe that the resource graph representation is efficient at describing the kinds of reasoning and learning that we observe in our physics education research studies. By choosing the appropriate scale of representation, we are better able to model the student classroom experience. Second, we believe that a representation which accurately accounts for reasoning in a given context is more likely to match the work done by curriculum developers whose task it is to affect student reasoning. Working at the micro- or macroscopic level of reasoning often does not match the learning goals of specific instructional materials. Finally, and most importantly, we find that an accurate representation lets us account for many different descriptions of student learning. We build on ideas such as phenomenological primitives,[1, 9] facets,[3] resources,[2, 4, 5, 12, 13] coordination classes,[9] and conceptual change theory[14-17] and find common ground with descriptions of specific student difficulties,[18] conceptual dynamics,[19-22] and certain kinds of analogical reasoning.[23] Our work is consistent with discussions of a "theoretical superstructure" to guide physics education research.[24] Our purpose is to help in creating a common language that is consistent with the community's work in research, curriculum development, and instruction.

The paper is split into five major sections. In section 2, we briefly review resources, coordination classes, and conceptual change theory. In section 3, we describe a simple representation of resource graphs, give examples of how resource graph can be used to represent student thinking, and discuss some shortcomings of our representation. Section 4 is the core of the paper. In it, we apply the resource graph representation to four types of conceptual change as defined by S. Demastes (Southerland).[17] Throughout this section, we compare our work to other models of reasoning and point out strengths and weaknesses of other approaches to modeling student thinking in the classroom. In Section 5, we apply the resource graph representation. First, we describe one form of analogical reasoning[23] and show that the resource graph representation can be applied to more than the four types of conceptual change. Second, we apply our representation to existing results from the PER literature to describe a kind of conceptual change not yet discussed in the conceptual change theory literature. The paper ends in section 6 with a discussion on the role of modeling in physics education research and a brief summary of our results.

# 2. LAYING THE FOUNDATION FOR THE RESOURCE GRAPH REPRESENTATION

We aim for brevity in reviewing the literature on resources, conceptual change theory, and coordination classes. We assume a reader's familiarity with the basic ideas of each model of reasoning and highlight the pieces we believe are most important.

## A. Knowledge pieces: resources

A growing body of evidence over several decades has shown that "knowledge pieces" are useful for modeling student reasoning. Pieces are small-scale knowledge elements that can be applied productively in many different settings.[3-10] For example of a knowledge piece, consider "part-for-whole," where a part of an object (or system) can represent the whole of the system.[25] This idea is natural in the context of politics (a president or queen represents a country), corporate culture (a charismatic Chief Executive Officer represents a company), and literature (as described by the term synecdoche). In physics, we use "part-for-whole" when we represent the motion of an object by the motion of its center of mass. In each case, our thinking is simplified by representing a large, complicated system by only a single part of the system: the center of mass of an object travels a parabolic trajectory, even as the object rotates, and so on.[1]

We may model many types of thinking – analytic, procedural, conceptual, epistemological, etc. – as being made up of such small-scale elements. Several examples have been described in the literature, including *agents*,[10] *phenomenological primitives*,[1, 9] *facets of knowledge*,[3] and *intuitive rules*.[8] For the purposes of this proposal we use Hammer's general term *resources*[4] to refer to any of these constructs. A classic example of a phenomenological primitive is "closer means stronger," which is true for heat by a fire, but applied incorrectly to explain summer warmth due to the Earth's proximity to the sun in summer.[27] Another example is "dying away,"[9] true for both the ring of a bell that has been struck and the motion a sliding box after receiving a strong push across a floor.

Resources are genetic and productive, meaning they motivate thinking and are the elements of thinking, as well. To describe thinking in terms of resources can focus our attention on what student *do* in the classroom rather than what they fail to do. It honors their actual thinking. On the other hand, modeling thinking in terms of individual resources may be too simple an approach to describe the richness of student reasoning.

## B. Classic, large-scale conceptual change theory

A different approach to modeling student learning has been suggested by researchers starting with Posner, Strike, Hewson, and Gertzog in 1982.[14] They built a model of conceptual change theory (which we refer to as "classic" conceptual change theory) based on Piagetian accommodation and the idea of the Kuhnian paradigm shift internalized into a single individual. The process of choosing between competing ideas was formalized by considering issues of plausibility, fruitfulness, and intelligibility of a new idea that might replace an old idea. An early assumption of rational consideration of competing models was later replaced by the possibility of the change occurring without the thinker's awareness. Modifications to the classic conceptual change model [28] included further (and possibly implicit) ways of comparing between competing models such as a refined description of the status of two conceptions.[15] Other discussions of conceptual change exist[29] and have guided research for several decades. New studies continue to be published on how to apply conceptual change theory in science[23, 30-33] and mathematics.[34-37]

---

[1] To go into more detail than will be given in this paper, the part-for-whole knowledge piece can be though of as a vital relation describing elements of a blend. For more information, see Ref. [26]. This paper will not discuss blends further, though they allow for an excellent, deeper, more detailed description of the use of knowledge pieces in reasoning.

Rather than summarize the history of conceptual change theory, we will focus on certain elements which are contained in the terminology of conceptual change theory and in ongoing research. First, from the original authors[14] forward, there has been an idea of a *conceptual ecology* in which the conceptual change occurs, that affects what kind of change occurs, and that provides the meaningful context for the change. Within this conceptual ecology, there are many different ways of knowing and understanding, including a person's epistemological stance, metacognitive beliefs, and connections to other, similar systems.[38] Several different types of conceptual change can occur, including *conceptual extension*, *conceptual exchange*, and more. We discuss these (using slightly different terminology) below.

### C. Coordination classes can change

One problem with classic conceptual change theory, as discussed by diSessa and Sherin,[11] is the lack of clarity in defining the thing that is actually changing in conceptual change. What, exactly, is a concept? How should one define the conceptual ecology? The authors propose that one type (out of many possible types) of concept is a *coordination class* consisting of *readout strategies* that organize sensory information and which activate a *causal net* of ideas that guide one's thinking in a given situation. DiSessa and Sherin show how "force" might be thought of as a coordination class and give examples from one student's development of the concept of force to show how much detail is necessary for a complete description of conceptual understanding. The idea of coordination classes has been applied to understand student reasoning in other areas, such as waves[25] and kinematics.[39]

Coordination classes are a way to create large-scale "concepts" based on a resources perspective. A network of resources is activated in a setting. For example, if an object falls toward the ground in front of a moving car, we may observe (read out) in particular that the object is small, orange, and moving slowly. It might be only a leaf. Reasoning resources (perhaps "small things have small effects" and "slow means soft") are primed and activated along with the readout. A combination of resources is coordinated (most likely subconsciously) to create a single behavior or result (for example, do not swerve to miss the leaf – but swerve to miss a small, grey, quickly falling stone). The model of resource coordination is consistent with what is known of long-term memory, namely that it is associative: activating one element "primes" other elements, making them ready for activation.

A detailed example illustrates the small-grain model. In previous research,[40] we found that student responses to a series of seemingly unrelated questions about mechanical waves could be explained as if students were thinking about physical objects, not propagating disturbances to a system. Students would read out that the peak of a wavepulse was much like the center-of-mass of an object; a readout guided by "part-for-whole" seemed to activate a set of ideas related to objects and a "causal net" of associated resources was activated. Students would then also often talk about effects "dying away," such as the amplitude of an ideal wave decaying during propagation. These students misinterpreted elements of the mathematical formulation of waves to arrive at their answer.[41] When discussing superposition,[42] students might activate resources such as "bouncing" and "canceling." Two waves traveling toward each other would not pass through each other but bounce off each other or part of one such wave might cancel another out permanently. When discussing propagation, students might apply "activating agency"[4, 9] (to quote a student, a "force of the hand" was needed to create the wave and get it moving, and a harder flick led to a faster wave[42]). When discussing sound waves propagating through the air, students might use "maintaining agency" (a force is required to keep things moving, for example), refined to include concepts of frequency and volume and their effect on the initial force on the system.[43] Each of these resources can naturally be applied to objects, but was applied to waves. We inferred the existence of an *object coordination class* being applied inappropriately in the context of waves.[25] The language allowed us to think of both the knowledge-in-pieces approach (resources)[1] and the conceptual approach (coordination classes)[11] and to tune our curriculum correspondingly[43] while remaining consistent with a description of specific student difficulties while learning new ideas.[40-42]

The coordination class model, in particular readout strategies and causal nets, is similar to other models of student reasoning. Readout strategies are akin to the term framing used by researchers such as Tannen,[44] Goffman,[45] and MacLachlan and Reid.[46] A frame is defined as the set of expectations about how to act and what to think in a given situation. Our discussion of readout strategies is tightly aligned with the description of framing given by Hammer, Redish, Elby, and Scherr.[47] Causal nets can be thought of in terms of semantic nets or the types of diagrams associated with connectionist theories.[48] We discuss the idea of linking resources in section III.

# 3. REPRESENTING COORDINATION CLASSES AS RESOURCE GRAPHS

In this section, we describe a simple representation of coordination classes that will allow us to discuss several meaningful aspects of the model. Our representation is incomplete but sufficient for our needs. Issues such as the structure of linking require further work.[49] We split our discussion into three pieces: an introduction to network diagrams, static and dynamic views of our diagrams, and a discussion of the flaws in our representation. In section IV, we will apply our representation to several different forms of conceptual change.

## A. Resource graphs – fractal networks of linked resources

We represent the resources that students use in a given setting by using a simple "circle and line" model. Resources (represented as circles) are linked together (connected by lines). We refer to our network diagrams as resource graphs. A simple version is shown in Figure 1.

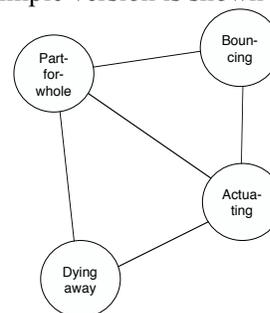

FIG. 1: Example resource graph. Individual resources are circled and connected to each other by lines.

### 1. Resources in networks

In our representation, each circle represents a resource. Resources, be they conceptual, procedural, or analytical, are represented as circles for the moment. In our figure, only conceptual resources are shown. We add simple names to help distinguish between resources. Only those resources used in a given situation are shown in our representation; one can imagine many other resources associated with objects, but our example shows only a few. Thus, our representation is a context-dependent resource graph that is at least theoretically observable. In the upper left of the figure, we show the perception driven conceptual resource "part-for-whole" which is connected to several other conceptual resources. The "net" of four resources describes one way of reasoning about a system as an object. You can throw objects with a certain speed, thus using an "actuating agency" to create motion. Objects bounce off other objects (which affects how they move, linking back to "actuating agency"). An object's motion dies away with time in a frictional world, which might be described as the actuating agency's effect "dying away." "Bouncing" and "dying away" are not linked in the resource graph because the two ideas are not used in conjunction with each other (though both are connected to "actuating agency," for example).

One can imagine experiments giving evidence of certain resources being used and activated in relation to others. For example, one can interpret data from Clement[50] to say that students use "actuating agency" and "dying away" to describe why a coin tossed into the air slows down until it reaches zero velocity at the top of its trajectory. That the coin flips in its motion is irrelevant; "part-for-whole" simplifies thinking so that the center of mass motion is the only relevant motion in the problem. Resources which are not observed (or whose traces cannot be inferred in a situation) are not included in the resource graph. In Clement's work, for example, there is no discussion of bouncing (though obviously coins bounce when they hit a table) and the above figure would contain too much information.

We note that our simple description of resource graphs is consistent with DiSessa's criterion of *span* for primitives and resources.[9] The span of a resource can be defined as whether or not it is applied in a given situation (e.g., whether it is part of the resource graph describing reasoning in that situation).

### 2. Activation and linking

Using one resource often primes another and makes the second more likely to be used in a given setting. When we observe or infer that the second resource has been activated, we say there is a link between the two. Links between resources are context dependent, consistent with our definition of resource graphs. We give many examples in a later section of this paper.

When the network of linked resources is consistently applied with strong linking between elements, it can be used to describe the *causal net* of a coordination class. We can imagine that one resource is first activated or used in a given context. Shown in Figure 2 is a typical readout when interpreting waves as objects: students often see the peak of a wavepulse and interpret it as if it were a center of mass of an object.[25] Other resources are activated subsequently. The activation of a resource and the subsequent linking structure can be compared to the *readout strategies* of coordination classes. Just as readout strategies depend on the causal net of available resources in determining how a situation is interpreted, the set of available resources in a resource graph influences which might be activated first in a given situation. For example, if "part-for-whole" were not available to the student at a given moment, then the observation of a pulse's peak would be unlikely to activate thoughts of an object. It is important to note that the linking structure of a resource graph depends on the first-activated resource. Had another resource been activated first (say, dying away in the figure below), a different set of links might have existed, though the resources in use would have stayed the same. (The last is by no means a necessary assumption but will not be discussed further in the paper).

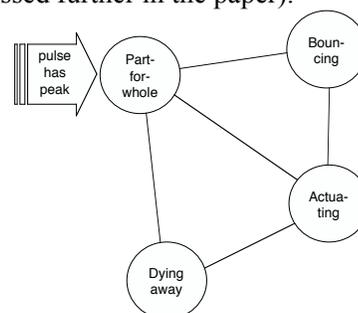

FIG. 2: Resource graph with activation included. This graph is built from data on student reasoning about wave-pulse propagation on a string.

### 3. Multiple scales for analyzing resources

Finally, our resource graphs allow us to consider multiple scales of analysis. For example, the resource "dying away" most likely contains much substructure. It describes a property of an object, where the property (such as the motion of a coin tossed in the air or the amplitude of a wavepulse traveling along a long, taut string) changes with time by slowly disappearing. Thus, objects, properties, and time dependencies all play a role in understanding the seemingly primitive "dying away." As a result, we can say that resources span across several levels of complexity, as described by Aufschnaiter and Aufschnaiter.[51] We represent this by showing sub-structure in a resource (see Figure 3).

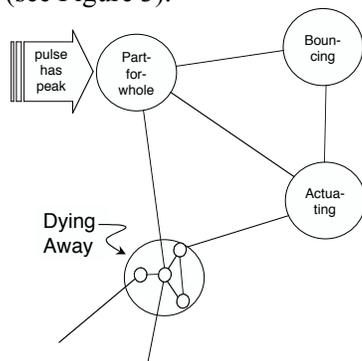

FIG. 3: Example of the fractal nature of resource graphs. The resource "dying away" is made up of sub-parts.

Several different readouts are associated with the observation of "dying away.," One needs to identify an object, look for information which describes the relevant properties, and use ways to see how the properties change with time. Such an analysis makes the resource "dying away" similar to a coordination class. We choose not to analyze resources at this level within this paper, though the fractal nature of resource graphs does suggest certain methods of interpretation which will be used in this paper. More details can be found in Sayre's work.[49]

In summary, resource graphs are representations of a person's resource use in a given situation. They are context dependent. They are consistent with terms such as span, readout strategies, and causal nets. They can describe the fractal nature of resources. At a basic level, they are consistent with the coordination class model.

## B. Examples of the dynamics of resource graphs

By choosing to represent resource graphs statically, we have left out several dynamic elements of reasoning which would require a dynamic representation. A movie perspective in which elements change over time or the path of activation can be followed might be more illustrative. We present two examples of how a dynamic resource graph might give insights into representing student reasoning. We leave further discussions for future research.

### 1. "Knowledge as invented stuff" promotes additional linking

In the first example, we consider activating one epistemological resource, "knowledge as invented stuff"[13]. When students apply this resource to their thinking, they might seek to create new ideas by combining existing conceptual resources into new constructs.[52] New ideas might be created along the way. The new ideas might activate previously un-primed ideas.

Representationally, activating the "knowledge as invented stuff" epistemological resource could have the effect of bringing more resources into a graph and increasing the likelihood of linking between them. Resources would gather on the graph and lines between resources would be created, perhaps kept for a while, and perhaps broken. Adding resources to a graph and increasing links between resources in a graph both describe student reasoning while creating new ideas. They are more free about considering ideas that at first seem foreign (*i.e.*, adding resources while brainstorming), and the ideas they use are often tested against other ideas more often (*i.e.*, linked to other ideas more firmly). That spirit of attempting something new is an observational measure of students using "knowledge as invented stuff."

### 2. "Knowledge as remembered stuff" suggests searching within a fixed net

In the second example, we consider activating a different epistemological resources, "knowledge as remembered stuff."[13] In such a situation, students may be trying to remember something they previously learned. They are not considering new connections between ideas, nor are they considering building new ideas. They are, instead, moving in a resource graph using existing linking structure, seeking an answer. For example, Tuminaro provides examples of recursive plug-and-chug epistemic games.[53] Often, such a method is appropriate and efficient. There are times, though, where the recursive plug-and-chug is inappropriate and new ideas are necessary in a way that the unchanging resource graph does not describe.

## C. Flaws in the representation

We have created a purposefully simple and easy representation of resource graphs. The reader may have raised many objections to the representation, two of which we wish to discuss before applying our representation to existing models of conceptual change and descriptions of learning in reform-based curricula. Many other objections can be raised to our representation, including the previously described static nature of the resource graphs.

### 1. Inadequate description of linking between resources

Our representation is unclear in showing whether a resource in a resource graph is actually used to reason in a context or whether it is primed for use but not yet activated. Using simple lines as links does not convey this information. Lines simply describe a connection between two resources.

Observation of resource linking is an experimental issue not discussed in this paper. From an observational standpoint, there is no difference between priming and activation; if resource use is not observed or inferred, then we have no evidence of it being primed, either. To address this limitation, we will assume that the resources graphs we draw are ideal graphs in which evidence for linking could be found in the correct experimental situations.

The links we show are drawn without blocks, stops, promotions, or other typical constructs of connectionist models.[48] Since the existing representation is sufficient to our needs of representing existing descriptions of conceptual change, we will not describe more complex link structures in this paper. Defining links (and finding appropriate observational tools to define them) is a major area of research when discussing causal nets and readout strategies.

### 2. Reconciliation and discrimination are not shown

The terms *reconciliation* and *discrimination* refer to two ways in which resources come to be seen as part of a coordination class.[11] We will not address these issues in this paper, nor does the representation include information about them. Instead, we will assume that the resources and links which are added to, deleted from, or modified when discussing resource graphs can be experimentally determined. Where possible, we will use existing literature to justify our claims, even when the original intent of the authors was not to give a resource-based description of student reasoning.

# 4. DESCRIBING CONCEPTUAL CHANGE USING RESOURCE GRAPHS

As stated above, several conceptual change theories have been proposed and discussed. We wish to express the process of change by looking at coordination classes and considering the changes that might be occurring to the coordination classes. We use our representation of resource graphs to represent the process of resource refinement, rearrangement, and restructuring which occurs in conceptual change.

To illustrate the applicability of our approach, we describe four types of conceptual change taken from Demastes:[17] *incremental*, *cascade*, *wholesale*, and *dual construction*. We describe two of these, incremental and cascade, as processes and two, wholesale and dual construction, as descriptions of states. For each of the four, we provide an example of the conceptual change from the existing physics education research literature. Where possible, we complement these examples with observations from our own teaching. We note that we do not rigorously show the existence of individual resources in these stories. Instead, we refer to existing descriptions of resources in the literature, and rely on plausible arguments at other times.

After applying resource graphs to describe the four types of conceptual change presented by Demastes, we apply our representation to two new areas. First, we describe one type of analogical reasoning in terms of the four first introduced by Demastes. We suggest that similar resource graphs allow for comparisons between similar situations. Second, we apply our representation to existing results from the PER literature to describe a new kind of conceptual change, *distinguishing*. In this form of conceptual change, a single resource graph (such as one connected to the idea of motion) is split in two as people learn to distinguish between velocity and acceleration.

## A. Incremental change as a process

### *1. Representation*

A common mechanism for learning is *incremental change*, in which a resource graph has resources added or deleted. The original resource graph on the left is modified in some fashion to have one new resource added to it, as shown in Figure 4. Adding a resource to a graph (meaning, linking this resource to others in a given context) requires that the span of the resource be increased to include the new situation. Other options might be creating a new link between resources, deleting a resource, or removing a link between resources. To delete a resource from a graph (*i.e.*, to change the causal net of an activated coordination class by removing an idea) requires that the span of a resource be changed such that it is no longer aligned with the situation. Incremental conceptual change is similar to Piagetian assimilation. Existing ideas remain, and new ideas are incorporated into the existing structure.

On purpose, we have left out several elements of our resource graph representations in our figure. We do not indicate the activation of the resource graph (equivalent to the readout strategy used to call up a causal net) because it is not important to this description. Also, we show only one form of incremental change. Other forms can easily be represented in similar figures.

We will use the representation of conceptual change, as shown above, in future examples. In this representation, "one concept" (on the left) is represented by a resource graph. It undergoes a change (represented by the arrow) so that at a later time it looks different in some specific fashion (as shown on the right).

### *2. Example from PER: Studying light from bulbs and shadows of masks*

Incremental conceptual change is common in traditional and reform-based instruction. The University of Washington (UW) *Tutorials in Introductory Physics*[54] activity on light and shadow,[55] for example, requires two incremental changes of students learning the material (see Figure 5). These incremental changes are isolated and do not depend on each other.

Students typically enter the activities with three ideas associated with the results of holding a mask in front of a bulb: light travels in straight lines, light travels out in all directions from a point source, and a mask that blocks light acts to create an inverted image of the light source.[56-58] In addition to difficulties in sketching the light coming through irregularly shaped masks, students are typically unable to answer questions about extended bulbs.[55]

The UW *Tutorial* on *Light and Shadow* first helps students strengthen their use of "light in a straight line" and "out in all directions." They sketch what is seen on a screen in the context of a simple point light source and a simple, symmetric mask hole and typically do very well.

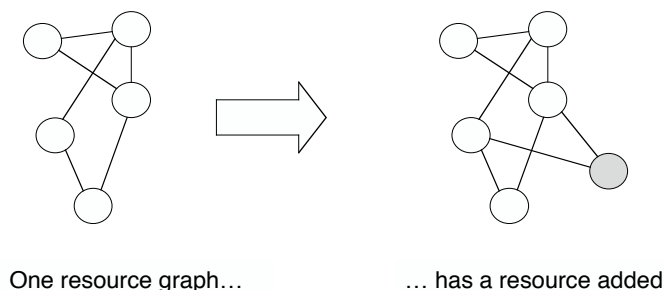

One resource graph…   … has a resource added

FIG. 4: Incremental change. A resource graph has a resource added (or taken away, not represented)

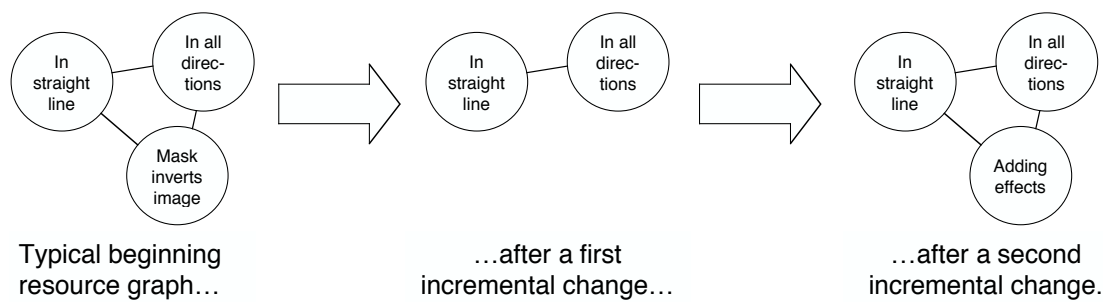

FIG. 5: Incremental change in studying light and shadow.
Two incremental changes occur, both deletion and addition

A first incremental change (of subtraction) is induced when students work through activities which prevent the use of "masks invert images" by considering irregularly shaped masks with light from point sources. When the image on the screen is not inverted, students must consider their model and perhaps decrease the span of "masks invert images" so that it no longer applies in this situation.

A second incremental change (of addition) occurs when "adding" is included in the resource network. Students must recognize that light travels out in all directions from each point on an extended source. They work with a long filament bulb and model it as a series of individual point sources. Each so-called point source creates its own shape on the screen behind the mask, and the sum of all effects is what we actually see. Thus, the "adding" (or superposition) resource is in play.[9,59] To activate "adding" is an incremental change to their existing network of resources.

In real-life instruction, both these incremental changes are often extremely difficult to carry out. Even when helping students explicitly develop the model of a long filament light bulb as a series of several small point-like light sources, we have found many students have a great difficulty incorporating the idea of "adding" in their reasoning.

Evidence for the disconnected nature of incremental changes comes from a recent examination in a recently taught course (using materials slightly modified from the original UW materials) at the University of Maine. We found that some students did not stop using "masks invert" and did not use "adding effects" in their responses. More importantly, some used both, in that they inverted the image due to the single point source and then added the effects due to each point source on an extended bulb (see Figure 6). Our finding supports the idea that incremental changes in resource graphs can be addition or subtraction from networks and that a series of incremental changes may be independent of each other.

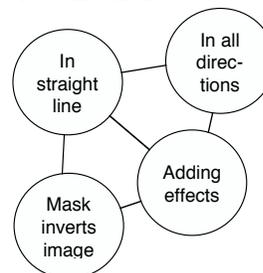

FIG. 6: Resource graph for incorrect response.
This student did not delete one resource but added another.

### 3. Other perspectives for describing incremental change

Other perspectives can be used to describe incremental changes, as well. From a Piagetian perspective,[60] a process of assimilation has occurred. The resource of "inversion" has been subtracted without causing major changes to the remaining ideas. The resource of "adding" has been assimilated into a larger structure, again without changing the existing ideas drastically.

From a resource perspective,[9] we see that the span of the "adding" resource has been expanded to include situations dealing with light and shadow. The resource is now aligned with this situation. Such a description is consistent with the description of conceptual change, but does not include a discussion of the ideas to which the resource gets linked.

Finally, from a classical conceptual change perspective,[14] we see that the idea of "inversion" has been found implausible (contradicting observations), while the idea of "adding" has been found fruitful, plausible, and intelligible. Classic conceptual change theory is an incomplete description because it does not fully recognize the ideas that remain in play and how they are part of a larger network of reasoning. Also, it does not account for those students who keep "inversion" while also using "adding."

## B. Cascade change as a process

Where a series of incremental changes can occur in isolated steps, it is also possible to have a series of incremental changes (addition or subtractions of resources from a network) occur in a connected fashion. These might be set off by a single change causing a cascade of further, connected changes (one can make an analogy to crystallization within a solution).

Though the term cascade change implies a quick series of events, it is impossible to know the speed with which they occurred. One might observe a student in the process of changing saying, "Oh! I thought it was X but now I see that it's Y, which means Z, as well. That's neat!" (We have observed such moments in a classroom, for example.) One could think of such an observation as evidence of building or restructuring a resource graph on-the-fly. But, cascade changes may occur in many other ways and need not be on-the-fly. One can imagine a slower, more methodical process of connected changes to a resource graph, in contrast to disconnected incremental changes.

### 1. Representation

At least two different types of cascade are possible, readout-cascade and network-cascade.

*Readout-cascade.* In the first, readouts that once activated a network or resources in a given setting get remapped to a new network through a surprising and far-reaching event. The effect can be called a 'cascade' in that it consists of a series of connected incremental changes, each dependent on the previous change. A readout-cascade is difficult to represent accurately in only two static resource graphs. In Figure 7, the shift is indicated by having the same activation now call up a new resource graph. The resources depicted in the graph on the right were not necessarily linked before the cascade change occurred (as represented by the new geometry of the system). A student might say, "Oh, if that's the case, then this other thing might also be true. Doesn't that mean that this third thing also occurs?" The full set of connected changes can be referred to as a cascade change.

*Network-cascade.* In a network-cascade, a resource within a network is changed in such a way as to create a cascading effect within the network as resources are dropped or added and new links are created between the resources. Again, the cascading changes are connected and dependent on one another. A single readout remains, but the core concept has

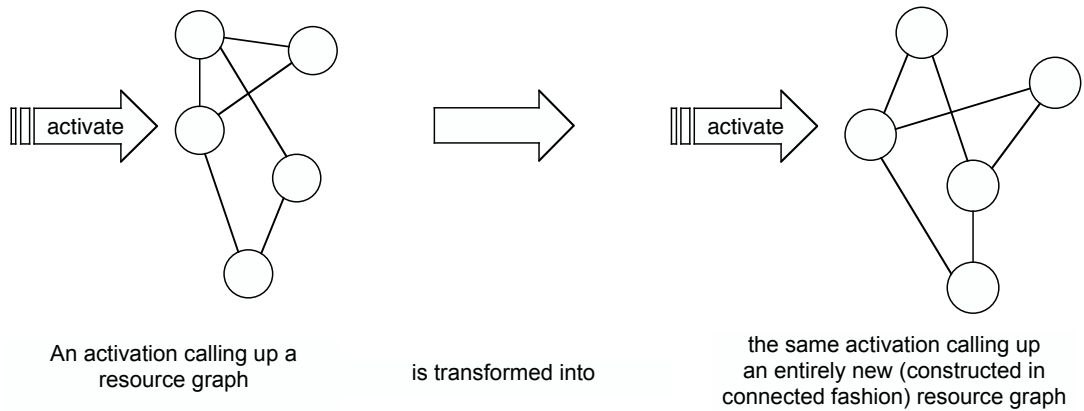

An activation calling up a resource graph    is transformed into    the same activation calling up an entirely new (constructed in connected fashion) resource graph changed.[17] In Figure 8, the activation and some of the resources (shaded grey on the right) remain the same while a cascade happens as one resource is suddenly connected to a whole set of new ideas. Again, the issue of on-the-fly construction cannot be addressed in the representation, though experimentally it might be observable. The resources that students use in their thinking would show up as new white circles in the right figure, but the time scale of how quickly students add them to their revised network of reasoning in a given situation cannot be represented.

FIG. 7: Readout-cascade change. The same activation leads to a new set of linked resources.

## 2. Example from PER

Cascade changes have been described in many physics education research settings. We have observed events (and have seen many presentations of similar events during talks at national American Association of Physics Teachers meetings) in which students struggle with an idea until suddenly an insight lets everything fall into place suddenly. We can paraphrase a typical student saying, "oh, I was thinking about it this way, but you want me to think about it this *other* way." At that point, a cascade of new ideas occurs as the student reasons about a situation and builds a new network of ideas.

Elby and Hammer [13] provide an example of readout-cascade change. By changing the readout of a situation, students can go through a cascade change which eventually leads to a cognitive conflict. Students are asked to reason about a collision between a truck and car. They quickly read out that two resources ("actuating agency" and "compensation") are in play. "Actuating agency" describes the effect of the truck colliding with the car and putting the car in motion. "Compensation" describes the response of the car to the collision; a heavier truck causes more effect. Students are first asked to apply these resources in the context of force. Compensation mediates actuating agency: students typically predict that the car feels twice the force that it exerts on the truck. Then, students are asked to connect the same two resources explicitly to changes in motion. Here, a cascade change occurs for those students who have not already connected "actuating agency" and "compensation" to acceleration. After reading out of the situation that "actuating agency" and "compensation" can be connected to changes in velocity, students can quickly and easily connect to ideas such as acceleration and Newton's Second Law, $F = ma$ and the Ohm's p-prim (in the context of $F = ma$).[9] Below, we describe as a dual construction what occurs afterward, when there are two different resource graphs which describe student reasoning in a given situation. The ensuing cognitive conflict can be used for teaching purposes.

Many other kinds of cascade changes occur in our teaching. A typical instructor tries to build a series of connected additions and subtractions to resource graphs and is typically unhappy when students make unconnected incremental changes to their thinking. We have often heard students go through a series of connected changes during moments of facilitation in small group learning environments. First one idea changes (either the

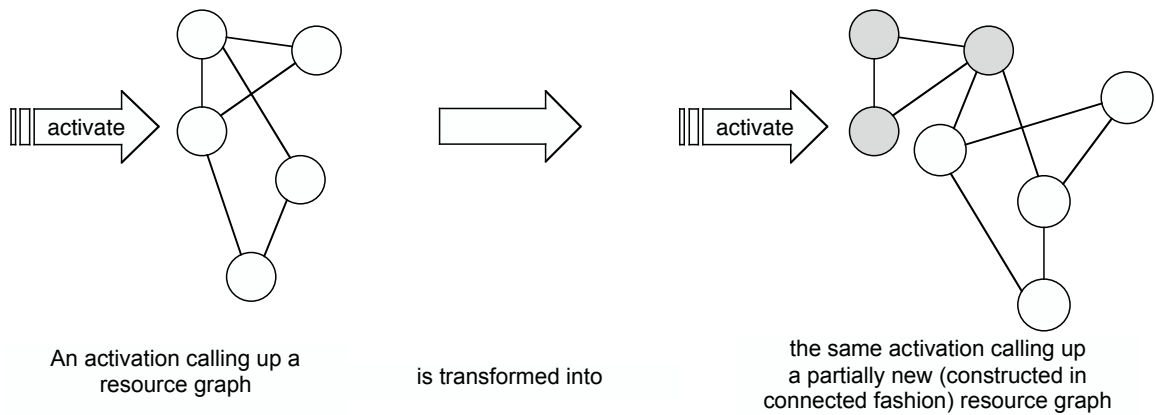

An activation calling up a resource graph    is transformed into    the same activation calling up a partially new (constructed in connected fashion) resource graph readout or a piece of a network in a resource graph); then, another idea gets evaluated and either changed, dropped, or incorporated; additional ideas get questioned more deeply, causing further changes; so on. Eventually, be it slowly over weeks or in the few minutes during which facilitation occurs, a whole new construct has emerged within the student's thinking.

FIG. 8: Network-cascade change. A change within a network leads to a series of connected changes.

### 3. Other perspectives for describing cascade change

Because cascade changes consist of additions and subtractions to resource graphs, the same perspectives that described incremental changes can be applied here. But, the same shortcomings listed above apply here, while the resource graph representation allows for a series of connected (cascade) or disconnected (incremental) additions and subtractions in ways that the other perspectives do not include.

One difficulty in comparing cascade change to other forms of conceptual change is that evidence of its existence relies on observations of connected changes. The observational task is substantial, and even the most detailed longitudinal study might not contain enough information. It is highly possible that many changes in students' reasoning about physics contexts are cascade changes which we do not observe. Even in small group work in a typical reformed curriculum, most students at a table are quiet. Thus, their learning cannot be observed using classroom video. In interviews, we may have more robust data, but people rarely publish entire interview transcripts and have not highlighted moments of cascade change in their presentations or publications.

## C. Wholesale change to describe states of graphs

Incremental and cascade changes describe processes which lead to a change in a resource graph. Others are also possible; we describe one, differentiation, below. We can also use resource graphs to describe different states of student reasoning without knowing the processes that led to the changes. Wholesale changes describe the changes from one resource graph to another over the course of time. The change may occur through incremental, cascade, or other as yet undefined processes.

### 1. Representation

When studying learning using pre- and post-instruction surveys, wholesale change can be observed for students who have moved from a common misconception to a correct expert model during instruction. In wholesale change, the final resource graph describing reasoning about a situation is wholly different from the original resource graph. Changes can occur with regard to readouts, the resources included in the graph, and the linking structure connecting the resources. Many wholesale changes can be shown, with the basic idea that the pre- and post-instruction resource graph look very different. Interim steps need not be observed.

To provide one example of a wholesale change, we represent students' resources being completely broken apart and unlinked only to be rebuilt in a new network. New ideas may be incorporated into an existing structure. Wholly new sets of readouts and reasoning resources may come into play. A period of profound confusion may exist in the interim (and only the most motivated students may persist in their studies along the way). This period of confusion is not necessary, but can be included in the representation to illustrate that intermediate steps are possible.

Figure 9 shows our process of wholesale conceptual change, including a possible period of profound confusion. A single resource graph is somehow transformed (through cognitive conflict or other processes) into a set of disconnected resources. Perhaps a student is "flailing about," trying to find the right idea, but the previous way of thinking about a situation no longer holds. Eventually, though, a new set of resources is built, with new linking structures. Some previous resources may still be useful in this new resource graph, while others may have been added to the network. For the sake of representational simplicity, we are not including activations, but it is highly likely that a changed network would lead to changed readout strategies.

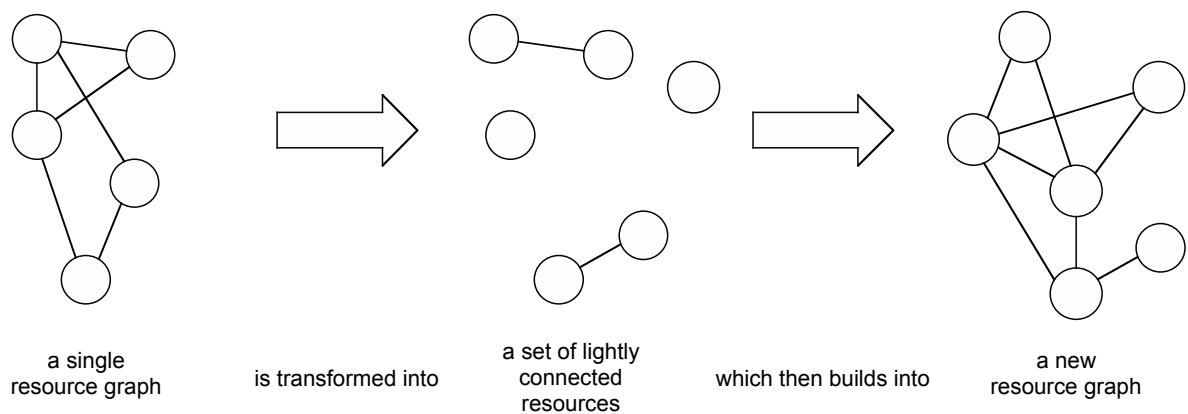

FIG. 9: Wholesale change represented in stages. At a later time, the resource graph describing a situation is entirely different from the original resource graph.

### 2. Example from PER

Examples of wholesale change can be found in the literature in situations where students move from one way of understanding the world to another, while keeping none of their original conception intact. One should not confuse a class's movement from 10% correct to 90% correct on a question as indicative of a wholesale change, since the model describes an individual's reasoning and not a class's collective behavior. Instead, one might characterize an individual student's responses to a well designed survey and characterize their understanding before and after instruction. For example, a student could move from an impetus model (where a coin tossed in the air "loses force" on its way up, as described previously in Clement's work[50]) to a Newtonian model on many questions of a survey such as the Force and Motion Concept Evaluation[61] Reif discusses such a change in terms of a transform matrix,[62] where there is a transformation of the student from naïve to expert learner. Bao and collaborators take a different approach, where they look for consistent models of reasoning before and after instruction rather than study student gains in correct responses.[63, 64]

It is possible to observe wholesale change in rare instances where one is observing students during the process. Such studies typically include sufficient interview evidence to characterize student reasoning before and after the moment of conceptual change. The example we present comes from work by Scherr, Vokos, Shaffer, and collaborators at the University of Washington. They could reliably create wholesale conceptual change in a surprisingly short time-scale when carrying out interviews on the topic of simultaneity in special relativity.[65, 66] In a typical interview task (described in full in their papers), two volcanoes erupt simultaneously in the reference frame of an observer on the ground. A spacecraft is flying with relativistic velocity from one volcano to the other. At the moment of eruption, it is directly over one volcano. Events 1 and 2 are defined as the eruptions of volcanoes 1 and 2, respectively. Students must decide whether event 1 happens before, after, or at the same time as event 2 for observers on the ground and in the spaceship. Scherr characterizes student reasoning[66] as made up of several knowledge pieces, including "visual reality" and "ultimate reality." Two instances of "visual reality" are found in student interviews. First, every observer has their own reference frame. Second, events are simultaneous if an observer receives light from the events at the same time. Scherr summarizes both these as "what you see is what there is." One of the many reasons why special relativity is counter-intuitive is that it violates the seemingly universal idea of "visual reality." Visual illusions may be surprising and amusing for the same reason. Students define "ultimate reality" in the simultaneity of events by believing that "things 'really happen' in only one way." This attitude may be correct for certain situations (rain falls downward) but not in others (many witnesses to a crime scene or accident scene may tell varied stories).

In Scherr's research, students often began the interview task by applying "visual reality" and "ultimate reality" in their answers. Thus, an observer on the ground equidistant from the volcanoes saw events 1 and 2 as simultaneous, but an observer standing at the foot of one and able to correct for light travel time would state that the events were not simultaneous. Through a precise set of interview questions (later turned into instructional materials), researchers were able to create a cognitive conflict in students. Students were confronted with a situation in which their use of "visual reality" or "ultimate reality" led to impossible situations. One could say that the resource graph describing a student's initial knowledge state fell apart. Scherr reports[66] that students experienced denial, withdrawal, or reached toward absurdist responses during the period of confusion. But, after a short while, they began to build a more accurate model of the situation in which appropriate ideas about simultaneity are used. Data suggest that the change from a single intervention is lasting and that students do not return to applying the ideas of "visual reality" or "ultimate reality" in similar contexts.

### 3. Other perspectives for describing wholesale change

Because of its prevalence as the measurable outcome of instruction, wholesale change is well studied and modeled. We discuss only three perspectives, though there are many more, including the process of model analysis mentioned above.[63, 64]

First is classic conceptual change theory[14, 15] in which Piagetian accommodation and aspects of Kuhnian paradigm shifts[67] are used to model changes in student reasoning. Because classic conceptual change theory includes the idea of Piagetian accommodation,[60] we address both at once. Instructors and researchers speak of replacing conceptions, consistent with the idea that the old resource graph no longer applies and a new resource graph is better able to describe reasoning in a situation. Where the original conceptual change theory includes a discussion of plausibility, fruitfulness, and intelligibility as a guide for determining what caused the one concept to be favored over another, the resource graph representation of wholesale change makes no claims about what causes the change in student reasoning. Many processes and causes for change are possible and need not be defined. Our representation implies that changes are driven by differences in activations caused by new readout strategies or modifications to the network of linked resources. In contrast, the Piagetian and classic conceptual change theories describe changes at a larger, concept level without going into detail about which elements of the concept change and which remain useful in a setting.

Another perspective which can be used to describe student reasoning is one of "overcoming a difficulty."[18] In keeping with the example from relativity, Scherr and her collaborators use the language of student difficulties when discussing the role of instruction on addressing certain aspects of student reasoning.[65] Students who have great difficulty with certain concepts permanently change their thinking in ways which can be consistently and repeatedly measured. But, as Scherr points out,[66] not all changes are lasting, and student reasoning must also be described in terms of changes to the individual resources and not just the whole network of ideas. Thus, the model of "overcoming a difficulty" does not completely describe student learning during wholesale change.

Finally, a process of wholesale change can occur when deciding between two models which are used to reason in a situation. We discuss such "dual constructions" below. For the moment, we can say there may be a period when one construction is useful to the student, but students learning a new idea come to value the second more than the first in the long run. A long time after the learning process has been completed, the change will seem like a wholesale change, even though the mechanisms involved were far more complicated. A process described by conceptual dynamics[19-22] can seem like wholesale change. At a long enough time scale, the original idea no longer holds and has been replaced by a new idea.

## D. Dual construction to describe states of graphs

In wholesale change, two different resource graphs exist to describe a single situation, but the graphs are separated in time. In contrast, it is possible to have two resource graphs to describe a single situation at the same time. For example, students might construct an idea in the classroom while reasonably holding on to the resource graph that they have used previously (and successfully) in their everyday life. Perhaps the most common form of conceptual change described in the physics education research literature is a process of eliciting dual constructions and creating cognitive conflict between them. We give two examples of dual constructions. One is an induced dual construction, where a second resource graph is elicited from students in addition to the first resource graph which they more readily use. Such a process implies building-on-the-fly, consistent with the idea of "building with," as described in the RBC (recognize, build with, and construct) model.[68] The other is a readout-based dual construction, where a student's readout of a situation determines which network is activated. A readout-based dual construction implies the existence of more fundamental, well formed, and lasting constructions.

### 1. Representation

The representation of dual constructions does not distinguish between induced and readout-based dual constructions. We will give an example in which identical readouts (for example, "which way does the object move?") with only minor variations ("am I in a physics classroom or not?") lead to drastically different responses. A classic question we have been asked during a quiz can be paraphrased as "do you want me to say what I think or what you taught me?" (Others have told similar stories, see ref. [69, 70].) A student asking such a question can apply (at least) two models to a single situation.

We represent the creation of multiple resource graphs in the figure below. A development is shown where a single network (activated in some fashion whose details are not important at the moment) changes such that the same activation now calls up two different, equally available and equally viable resource graphs. For pictorial simplicity, we do not show that these networks share elements, though in reality they may share many elements. In the case of the classroom question about motion described above, the activation may be a simple physics question. The one network might be the everyday language a student brings to the classroom and the other might be the physics language learned during the course. In general, the dually activated resource graphs have no inherent rightness or wrongness associated with them. Where in one class we wish students to apply more formal descriptions of a situation and not apply their colloquial language (each represented by a different resource graph), other situations such as wave-particle duality require correct and at times contradictory dual constructions. An example of necessary dual constructions can also be found in the application of band structure and Drude (semi-classical electron) models of conductivity.[71]

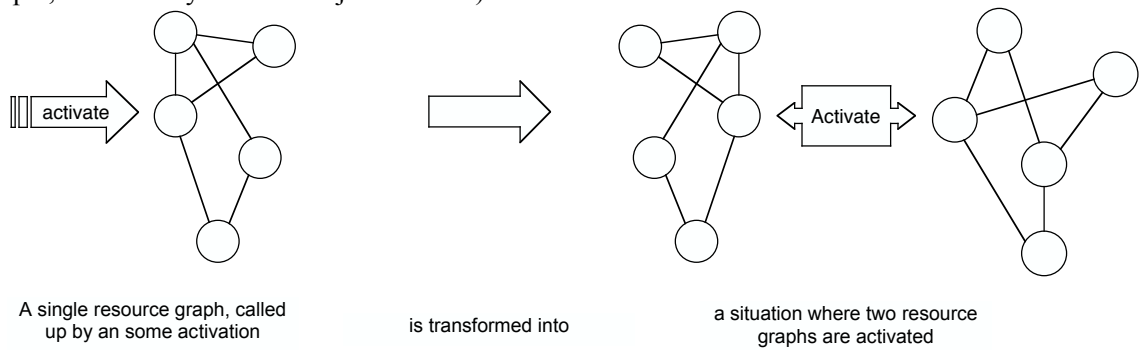

FIG. 10: Dual construction change. In a single situation, one is able to reason clearly using two different resource graphs.

### 2. Examples from PER

We illustrate the richness and prevalence of dual construction conceptual change by giving two examples from the PER literature. The first, in which two models of work are presented to students and they must learn to choose which is important, comes from the UW *Tutorials in Introductory Physics*.[54] The second, in which both force and acceleration are triggered when considering a collision between objects, comes from the University of Maryland *Learning How to Learn* project[13, 72] and has already been described above as showing evidence of a cascade change.

*Readout-induced dual constructions.* While studying student understanding of the work-energy and impulse-momentum theorems,[73, 74] researchers at UW created situations involving different mass objects pushed by constant (and identical) forces across essentially frictionless surfaces. In such a situation, applying $KE = 1/2\ mv^2$ and $p = mv$ to understand changes in kinetic energy and momentum, respectively, lead to inconsistent responses, while applying $\Delta KE = F\Delta x$ and $\Delta p = F\Delta t$ gives consistently correct answers.

In interviews and on written responses, students typically spoke or wrote in terms of velocities and masses. The readout of the situation (observing masses moving) reasonably led to mass and motion-based reasoning. To arrive at the desired comparisons, students typically carried out compensation arguments: the larger mass accelerates less, but for a longer time, so it has the same kinetic energy than the smaller mass in the end. The same compensation argument might lead to the conclusion that the final momenta are also the same. Both answers cannot be true.

A different readout of the situation was available to students, if they noted that the force exerted on the masses and the distances they traveled were the same. In interviews, students who did not think of this readout of the situation were explicitly reminded of it.

What we see is that students can use two arguments to consider changes in kinetic energy, one based on a readout of masses and motion, one based on a readout of forces and distance. Given certain assumptions, answers might seem identical for students. But, when considering additional evidence (or a different quantity), the dual construction leads to conflict. From this, researchers and curriculum developers at the University of Washington created instructional materials[54] shown to be successful in helping students learn the physics.[74] Based on the resource graph representation, it is clear that the issue is not whether students can reason with force, distance, and time, but whether they activate that network of ideas over another, possibly more compelling network (perhaps because it involves changing quantities that are more visually compelling[75]).

*A need to refine raw intuitions.* The Newton's Third Law tutorial by Elby and Hammer has been described above as containing a cascade change in which students apply two resources to the context of velocity change and then reason through several steps to arrive at conclusions about acceleration and the force that two objects exert on each other.[13] Prior to this, students have also applied the same two resources in the context of force. When asked to compare the forces a car and a truck exert on each other, students, even those who might know the right answer, will say "well, of course you would *think* it's going to be the truck." A dual construction has been induced in the students. One situation calls us two different resource graphs (though they share two resources, "actuating agency" and "compensation") which lead to conflicting descriptions of the forces a car and truck exert on each other. Discussing the conflict forms the core of Elby's instructional materials. As described by Elby and Hammer, the students are applying an intuition of "feels more effect" in two different settings, acceleration and force. The cognitive conflict that is created comes from a dual construction induced by a cascade change effected in the students; the authors actively help the students build two different networks of ideas that cannot both be true.

### 3. Other perspectives for describing dual constructions

Many other perspectives have been used to describe dual constructions, including classic conceptual change theory, hybrid and mixed state mental models, conceptual dynamics, and situated cognition.

In classic conceptual change theory,[14] researchers discuss a student's shift from one conception to another. There is little discussion of how the two conceptions came to be, but there are several criteria that determine the status of one conception compared to another. A new idea must be intelligible, fruitful, and plausible. In our language, a resource graph must be created for this new idea, and the results of thinking using the ideas contained in the resource graph should be valuable to the student. We believe that there is rarely sufficient evidence to describe a student's conception. Instead, we believe that a smaller scale construction of several resources more accurately reflects student reasoning about competing ideas.

Some researchers describe students who "have" two different theories when answering a set of questions. Hrepic describes hybrid states of mental models.[76] Students are considered to have multiple complete and robust models of a situation, and these compete with each other in a given situation, so that only one or the other is used on a given question. Bao and Redish discuss a similar situation in terms of mixed mental model states.[63] The data discussed by these researchers show that students construct different resource networks for a given situation. We do not believe it is appropriate to make claims about students' conceptions when dealing with individual questions, and believe a model based on networks of resources more accurately describes our observations of the fluidity of student reasoning. It may be that the description of hybrid and mixed mental models in students' minds is an artifact arising from comparisons between students responses and experts' formulations of models which are consistent with, but far more detailed than, what students say.

One can speak of the conceptual dynamics,[19, 20] for example when there is a transition from an "old" to a "new" point of view when studying kinematics.[21, 22] As with hybrid models, there is both an assumption of information for which there might not be evidence and a disconnect between the fluid reasoning we observe in many of our students and the robust nature of the "points of view" or "concepts" whose changes are being studied.

Finally, one can think of situated cognition[77] as a way to describe what happens with student reasoning as they construct multiple, situation- and context-dependent ways of thinking about a situation. Though consistent with the large-scale readout of "in a physics classroom" or "in the non-classroom world," situated cognition does not provide an explanation sufficiently detailed to account for the subtle ways in which readouts of mass and motion are chosen over readouts of force and distance in the work-energy theorem example given above. In light of the subtleties of reasoning evident in many physics education research studies, the dual construction resource graph provides a better match to data gathered in the classroom.

### E. The value of resource graphs in applying conceptual change

Our goal in this section has been to illustrate the value of choosing appropriate models in physics education research. One's choice of representation can play a profound role in guiding our research, instruction, and curriculum development. We argue that most researchers implicitly use a "mescoscopic" model of student thinking, somewhere between individual, fleeting resource activation and the replacement of robust, large-scale concepts. We believe that our language must be brought into better alignment with the work that we do. After describing two forms of conceptual change using the resource graph representation, we suggest experimental work which is required to better understand the process of learning in the various reform-based curricula.

The mesoscopic scale allows us to be more consistent with the kinds of activities we see in students and in researchers studying students. None of the four types of conceptual change described above actually involved "concepts," in the sense of large-scale, robust, coherent structures of student reasoning. DiSessa and Sherin state that coordination classes are one possible kind of concept. The data they use [11] to prove the existence of a coordination class (in their case, "force") is very detailed. Most studies in PER are not as careful and complete. From the (mis)conceptions perspective, researchers rarely present the data required to convincingly show that students have learned a concept (or that one concept has replaced another). From the resources perspective, one can discuss whether individual resources are used appropriately or not in a given setting, but researchers are aware that reasoning is far more complex than whether an individual idea is turned on or off in binary fashion. The mesoscopic resource graph representation of coordinated sets of ideas allows for evidence-driven discussion of complex student reasoning.

A mesoscopic model should take on characteristics of both micro- and macroscopic systems. We believe that the resource graph representation can be used to show the large space of reasoning in which ideas are neither fluidly, fleetingly held (as resources often are) nor robust and permanently stored (as concepts are). Most student learning occurs where ideas are neither. Students are learning new ideas, incorporating them into their thinking, breaking apart connections and rebuilding them in new ways, compiling ideas into new structures, and taking these compiled structures into new situations. They are constantly evaluating (or being evaluated) on what they are learning, struggling to build whole new networks, and often struggling to understand the *kinds* of networks which we use in physics.

The four different types of conceptual change theory described above differ in several ways, including: their role in instruction, the time scale required for the change, and the experimental ease of observing any conceptual change. Furthermore, from the descriptions given above, it is clear that the types of conceptual change are not exclusive. Wholesale change might involve incremental changes leading to dual constructions which eventually cascade into a single, new construct. Other combinations have also been described above. There may be subtleties involved in how typical students progress through certain curricula. In some cases, such as when learning about dynamics in a physics class, one may wish to prevent the creation of dual constructions and have students use an epistemological stance of coherence[4, 13] at all stages in their learning. In other cases, such as when learning wave-particle duality, one may need to insist on building dual constructions before one can move on to merging the ideas into a coherent model of quantum physics. Similar care might be taken when considering how to induce incremental or cascade changes in instruction.

## 5. APPLYING RESOURCE GRAPHS TO OTHER FORMS OF CONCEPTUAL CHANGE

Having described four types of conceptual change in terms of resource graphs, we wish to extend our description to two other situations. This section of the paper shows only two applications of the resource graph representation as a way of binding together many different results from physics education research. Many other applications are possible.

### A. Describing one form of analogical reasoning

Analogical reasoning consists of comparing one situation to another and using the rules of one situation to guide thinking about unknown rules in another situation. For example, Duit[78] creates teaching situations in which analogies are used to create conceptual change in students reasoning about chaos. He finds that appropriate connections between two systems can help students' conceptions of an idea change as they use the analogy to gain new ideas. Two elements of Duit's work lend themselves very well to a resource graph representation.

First, we can describe the analogy itself in terms of resource graph. We describe the analogy as a comparison between two very similar resource graphs. Some elements are shared or at least very similar, and the links between these elements are identical (or nearly so) in both situations. One situation, the target analogy, has a more complete resource graph than the other. In applying the analogy, one can use the more complete resource graph to ascribe properties and links to resources not yet associated with the less complete resource graph. In an example not from Duit's study, one can consider water flow as a way to model circuits. One can think of flow down many parallel channels as being a shared property of water flow and electric circuits. For students who have not thought about the issue of charge conservation, the analogy to water flow can act as a natural guide in reasoning. Thus, the "conservation" resource (perhaps thought of as "nothing gets used up") can be brought into reasoning about electric circuits for those students who had previously thought of current being "used up."

Second, we can describe the conceptual change that is created in analogical reasoning. It seems to be an incremental or cascade change. The change cannot be wholesale because the analogy remains; the similarity of resource graphs remaining prevents one of the graphs from changing in a wholesale fashion. The change might be cascade, in that the changes in resources might be connected together as they develop due to the analogy. The result of the change might be the shift of a dual construction shifting to a single view of a situation; by strengthening the analogy, one might help students stop using one method of thinking about a situation when they previously had multiple ways. Several other possibilities can also be considered.

In closing, we point out that analogical reasoning as described from a resource graph representation is a productive way to use the language of conceptual blending[26] in existing models of physics education research.

## B. A new kind of conceptual change: Differentiation

We have so far shown that five types of conceptual change can be described by resource graphs and are consistent with different kinds of learning shown in the physics education research literature. Rather than describe existing conceptual change theories in the context of existing PER results, we can use PER results to describe a type of conceptual change not yet discussed in the conceptual change literature. Rather than creating a dual construction in a single context, *differentiation* is the process of learning to see two situations in what was once a single context.

We introduce the idea of *differentiation* using research by Dewey Dykstra and others on student learning of kinematics to describe a process of differentiation between velocity and acceleration. To account for findings, we introduce a representation of differentiation which we then apply to student learning of quantum tunneling. Data are taken from data gathered at the University of Maine by Morgan, where students learn to differentiate between energy and probability in the case of tunneled particles.

### 1. Examples from PER: Motion is the conflation of velocity and acceleration

Rather than begin with an idealized schematic to represent student reasoning, we first discuss an example from the PER literature as a way of motivating the need for the idea of differentiation. Typically, students enter our courses conflating the ideas of velocity and acceleration into a single description of motion.[21, 22] Acceleration is only an increase in velocity; deceleration, a decrease in velocity, is a different idea. Dykstra describes that these ideas are combined into a general concept of motion. His work goes on to discuss ways of separating motion into two concepts, velocity and acceleration.

Supporting evidence for the conflation of velocity and acceleration in motion comes from several sources. Questions on the Force and Motion Concept Evaluation (FMCE)[61] allow researchers to see how students view acceleration. On the coin toss question described earlier in this paper,[50] some students say the acceleration of a tossed coin points in the direction of the velocity on the way up and the way down, and is zero at the top of the trajectory.[61, 79] Pre- and post-instruction data from the FMCE supports that students move from a conflated view of motion to a differentiated view of velocity and acceleration. Further evidence comes from Shaffer's work,[80] where students show a confusion in their use of vectors describing velocity and acceleration. After targeted instruction, students no longer confuse the ideas. Again, they have learned to differentiate between velocity and acceleration.

As Dykstra points out, helping students see motion as two different concepts, velocity and acceleration, is central to teaching students to understand force and Newton's Laws.[21, 22]

### 2. Representation

To account for differentiation in ideas, we introduce a simple resources graph that shows differentiation in the figure below. We begin with a relatively large resource graph with many elements that are only loosely connected and generally activated by an observation (for example, the motion of an object). This graph is transformed into two distinct resource graphs, each with its own activation. The term differentiation is chosen specifically to describe the closeness of the two resource graphs and that one must seek out subtle differences in order to appropriately differentiate between two related networks of ideas.

For the moment, we do not distinguish between types of differentiation in our resource graph representation, though several might exist. We give one example below. Many others exist but are not discussed here.

### 3. Differentiation in quantum tunneling

In work carried out at the UMaine Physics Education Research Laboratory, building on work begun at the University of Maryland,[81, 82] we have found that students often think that both energy and probability of finding a particle are lost when quantum particles tunnel through a barrier.[83-85] A typical question involved students discussing the energy of particles which tunnel through a finite square barrier, with the particle energy less than the height of the barrier. A large number of students say that energy is lost by the particles which have tunneled. Also, nearly all students correctly state that there is less probability of finding a tunneled particle than finding one that has not tunneled. We can describe the students as having one large resource graph, activated by the observation of a barrier, including such resources as "maintaining agency," "dying away," "overcoming," visual markers such as "height of a graph," and conceptual markers such as "number of particles." Other resources can be connected, such as "constant" for the idea that some things remain constant (perhaps the energy of the incoming particles) as seen by the constant wavelength of the particle's wavefunction. The resources of "dying away" and "overcoming" are applied to both energy and probability in student reasoning.[84, 86] Figure 12 shows a typical tunneling scenario with a square barrier and Figure 13 represents how students come to differentiate between the different resource networks.

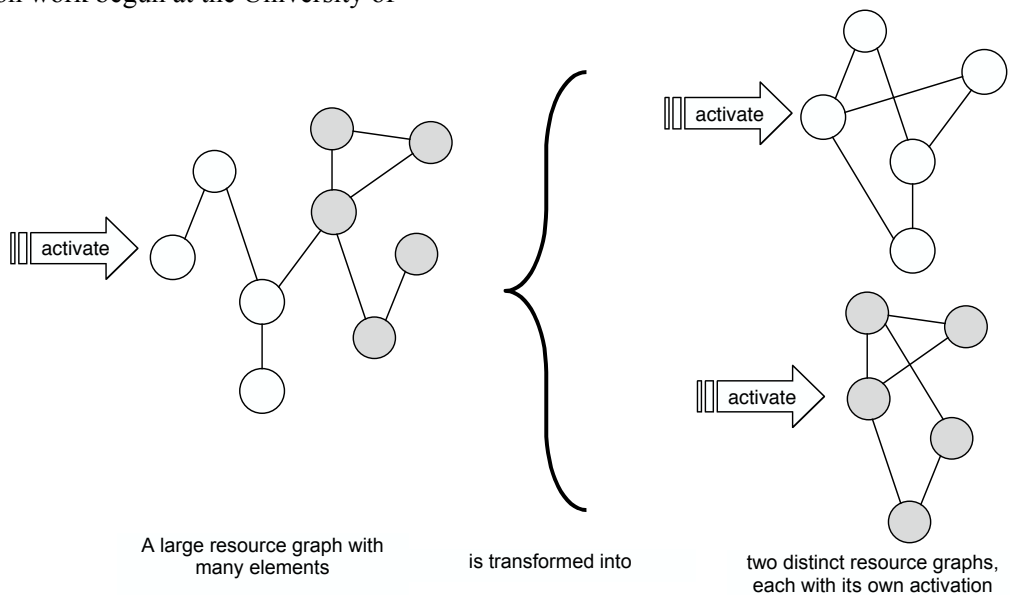

A large resource graph with many elements　　is transformed into　　two distinct resource graphs, each with its own activation

FIG. 11: Differentiation as conceptual change. A previously connected set of ideas is separated into two resource graphs each with its own activation.

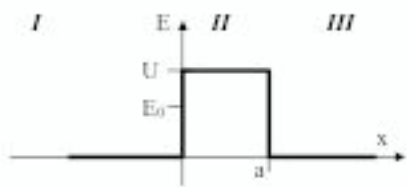

FIG. 12: Typical tunneling problem. A beam of particles with energy $E_0$ is incident on a square barrier of energy U. Regions I, II, and III are labels.

In Figure 14, we show an example of a typical student response when conflating the different elements of the preliminary resource graph. Note that the "visual: height" resource is used to indicate the lost energy of the particles that have tunneled; the axis around which the wave function oscillates is lower than before.[85] Note also the inconsistency in the student's sketch: the wave number, $k$, dependent on particle energy, is the same in every equation the student writes, as is the wavelength of the wave function. Yet, the student states "[the particles] in region 3 'lost' energy while tunneling through the barrier." The student sketches the wavelength as constant, but connects the idea of energy to the visual cue of the height of the oscillation axis to the energy of the particle. His understanding is best described by the resource graph on the left of Figure 13, where both energy and number of particles that tunnel are connected to "dying away" and "overcoming."

We have found that students who correctly learn the physics of quantum tunneling learn to hold to a constant energy throughout the entire physical system, shown on the right in Figure 13. They must learn to draw wave functions differently, always oscillating about the axis. They can then more easily recognize that the wavelength (and their mathematical formalism) already tell them about the energy of tunneled particles. The amplitude of the wave function may shift, but that is consistent with the assumed decrease. Something still "dies away," after all. In summary, students learn to differentiate between previously aligned resources in order to create two separate resource graphs, activated by different readout strategies.

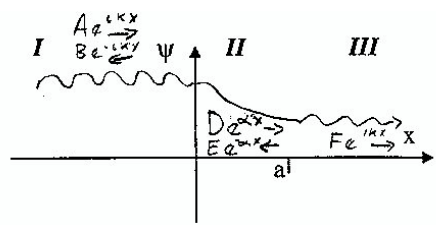

FIG. 14: Student response to tunneling question. A mixture of correct and incorrect ideas is applied in contradictory fashion. Differentiation has not yet occurred.

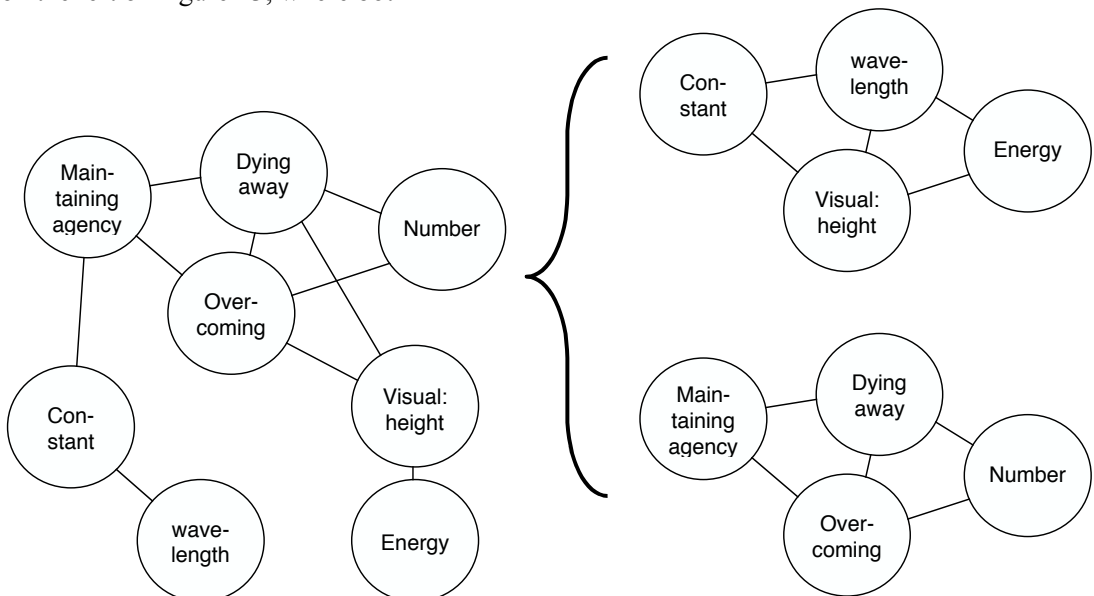

FIG. 13: Differentiation when learning quantum tunneling. A previously connected resource graph is split into two different graphs, one dealing with energy and one with number of transmitted particles.

Differentiation as a form of conceptual change is common in physics learning. Examples exist when learning motion (velocity and acceleration), quantum tunneling (energy and probability), energy and momentum (as described by Lawson[73] and O'Brien Pride[74]), and areas where students have difficulty differentiating between a quantity and its rate of change. Still, it has not been discussed in the conceptual change literature previously, showing that the resource graph representation when applied to physics education research results leads naturally to extensions of theories into new areas.

## C. Further applications of resource graphs

Other examples of conceptual changes easily represented by resource graphs are also possible. We suggest, but do not present evidence for, changes such as compilation and re-framing. We also suggest studies which would help elucidate the details of cascade changes as a way of distinguishing between curricula which teach Newton's Second or Newton's Third Law.

### 1. Conceptual change: Compilation as a process

Compilation might describe how larger scale concepts come to be as resources combine fractally into larger, more solidly linked structures. Resources combine into such firmly connected structures that they begin to act as individual resources (with no functional individual elements) in larger structures. In ongoing work at the University of Maine, Eleanor Sayre is leading a project to understand the process by which resources come to be. She is studying how fluid ideas become more plastic and then solid with time, how ideas combine, and how resource graphs reify into larger structures. She has describe a scale of "plasticity" to account for the different levels of commitment and duration of knowledge as students develop new resources.

### 2. Conceptual change: Re-framing as a process

Re-framing might describe how activations in a resource graph are changed. By changing one's activation of a graph, different connections might be promoted and strengthened, leading to nuances in one's reasoning about a situation. In another project at the University of Maine, Padraic Springuel is studying how visualization affects student reasoning about concepts such as kinematics and integration. Preliminary results by Thompson (on which Springuel is building) show that students re-frame situations using subtle cues to incorporate resources in one setting that are not part of another, seemingly identical setting.[87]

### 3. Understanding differences between curricula

When introducing Newton's second law, one can begin by considering either a single force or many forces acting on a mass. The former causes acceleration, which is difficult for many students to understand. The second requires vector sums, which can be differently difficult to understand. Two leading reform-based curricula approach the problem from each viewpoint. In *RealTime Physics*, students work with low friction carts and a mass hanging from a pulley to accelerate a series of differently massed systems. They develop Newton's Second Law from this single force and only then move to a description of competing (and sometimes equal in magnitude) forces. We can imagine this process as a cascade change, for example: starting with the observation of carts speeding up, "causation" linking the speeding up to the mass hanging from the pulley, "Ohm's p-prim" accounting for the ratio of cart mass to acceleration, and then "balancing" for competing forces that happen to be equal. In the *Tutorials in Introductory Physics*, students first discuss a stationary object on which many forces are acting. They learn to distinguish between contact and non-contact forces, compare systems in horizontal and vertical systems, and generally use a static situation to derive the idea of Newton's Second Law. We can imagine this process as using a very different set of resources, for example: including "balancing" from the very beginning, and incorporating resources related to Newton's Third Law much sooner. The resource graph that is built by the students might differ substantially from that developed in *RealTime Physics*, yet both are correct resource graphs for describing Newton's Second Law. Which is best for students might depend on the ease with which students link the specific resources in each form of cascade change.

Other experimental studies can also be suggested to test how students' conceptions change over time. In preliminary work carried out by Trevor Smith at the University of Maine, we have compared three ways of teaching Newton's Third Law, using the tutorial by Elby and Hammer described above,[4, 13] the University of Washington *Tutorials in Introductory Physics*,[54] and the MBL-based *Activity-Based Tutorials*.[88] In two years of instruction, we have found that students using the tutorial developed by Elby consistently perform significantly better on a variety of examination and standardized test results (using the FMCE[61]) in both pushing and collision situations that require Newton's Third Law to be analyzed properly.[89] At one level, our data are satisfying, since we can say that for our institution there is sufficient evidence to show that one method is more effective than another. At another level, our result tells us far too little to help others make adequate instructional choices. Without observational classroom data, we do not know which tutorial best matched lecture instruction, which questions in the tutorials were most effective in creating effective learning in students, or how individual questions affected students and caused a change in their understanding. Describing our results in detail, using ideas of resource activation and representations of the resources graphs developed in each instructional setting, would strengthen our understanding of the actual learning processes promoted by each curriculum.

# 6. SUMMARY

We have described several types of conceptual change using resource graphs to build a mescoscopic description of linked resource use. Our representation of resource-based reasoning that can account for several seminal results in the physics education research literature. The representation consists of a graph showing sets of linked resources activated in a specific context. Changes to the representation can involve resources being added or dropped from the graph, links being dropped or added, and changes in the activation of a set of linked resources. Our data has been connected to experimental work carried out at many different institutions over several decades. Our goal has been to provide a language that is relatively simple to access and better matches actual events in student learning and curriculum development than either the small scale, microscopic resources model or the large-scale, macroscopic misconceptions model.

Using the resource graph representation, we are able to describe several published forms of conceptual change and discuss several other types. In incremental change, a single resource is added or dropped from a resource graph while leaving the rest of the graph largely intact. In cascade change, a small change in one area can cause a large-scale shift in the resource graph as connected changes cascade through the system. In dual construction conceptual change, one builds multiple resource graphs that are activated in parallel. Finally, in wholesale conceptual change, the entire resource graph changes. We use our representation to describe analogical reasoning as a mapping of one resource graph onto another. We introduce the term differentiation to describe the conceptual change that occurs as students learn to create two distinct resource graphs out of one large graph that described closely related ideas.

It is clear that the many types of conceptual change are not exclusive. Some of the conceptual changes described in the literature are processes, while other describe differences between states in students minds. Wholesale change might involve incremental changes leading to dual constructions which eventually cascade into a single, new construct. Other combinations have also been described above. There may be subtleties involved in how typical students progress through certain curricula. In some cases, such as when learning about dynamics in a physics class, one may wish to prevent the creation of dual constructions and have students use an epistemological stance of coherence[4, 13] at all stages in their learning. In other cases, such as when learning about conductivity, one may need to insist on building dual constructions before one can move on to merging the ideas into a coherent model of quantum physics.[71] Similar care might be taken when considering how to induce incremental or cascade changes in instruction.

We argue that the resource graph representation allows a succinct description of learning on the scale that students experience in our physics classes. Students typically struggle to learn new ideas piece by piece but not entirely in isolation. Their learning often comes in isolated chunks which must be combined into larger chunks later on.[51] Many ideas are taught on the assumption of incremental changes or cascade changes which build from previous ideas. Typically, students address dual constructions in a specific context such as Newton's Second Law, and the results should not be assumed to be universal as the students are often not explicitly taught to generalize results across situations. At times, we see a fundamental change in how students think about an entire topic, but wholesale changes are rare on the time scales in which we can observe students and convincingly show that a complete change in reasoning has occurred.

Our representation does not include information about how to create the types of conceptual change in a given form of instruction. Incremental changes seem simple and quick yet can be very difficult to create and may take a long time to cause in a student. Cascade changes are often the desired pathway for building a series of connected steps in learning a new idea but may be similarly difficult to create. Dual constructions may be a natural (and negative!) way of separating classroom learning from everyday thinking but can be induced in the short term through appropriate activities. Wholesale changes may or may not involve a period of confusion while leading to one idea being replaced by another. Differentiation might be a natural element of instruction but not made sufficiently clear to the students.

Other models of reasoning can be accounted for using the resource graph representation. Conceptual dynamics in which there is a shift from old (typically incorrect or incomplete) to new (and aligned with experts) views might be thought of as the time development of dual constructions leading to a wholesale change in students. Specific student difficulties in which students fail to apply some element of thinking can be thought of as incomplete resource graphs, a false application of a resource graph, or simply a mapping of what resources are and are not part of the context-specific resource graph that students bring to a situation. Broadly speaking, Piagetian assimilation can be thought of as incremental change and Piagetian accommodation can be thought of as wholesale change. Analogical reasoning can be thought of as the comparison of similar resource graphs, one more advanced than the other, and one's use of similarities between the two to create incremental changes in the less complete graph. Finally, our representation of resource graphs is consistent with diSessa and Sherin's coordination classes model but extends it to mesoscopic grain sizes for which experimental data can be less rigorous, changes in reasoning can be more fleeting, and the structures being described are much smaller than concepts such as "force"[11] or "object."[25]

The resource graph representation not only accounts for results from the existing physics education research literature but suggests new experiments and the need for additional data sources within existing experiments. Creating an appropriate representation helps us clarify our thinking, seek new results, refine our experimental work, and discuss results with a consistent language. The resource graph representation, though incomplete, moves us forward in each of these directions.

# Acknowledgments

The ideas in this paper were first presented at the January, 2003, AAPT National Meeting in Austin, TX,[90] and further expanded on at the August, 2004, AAPT National Meeting in Sacramento, CA.[91] The work in this paper would not have been possible without discussions with (in alphabetical order) Andrea

A. Disessa, John E. Donovan II, Andrew Elby, David Hammer, Michael A. Murphy, Jeffrey Owen, Jonathan Pratt, Edward F. Redish, Eleanor C. Sayre, Rachel E. Scherr, Bruce Sherin, R. Padraic Springuel, John R. Thompson, and Adrienne Traxler. I indebted to the many years of feedback provided by this group. This work was supported in part by NSF grants DUE-0410895, DUE9652877, and DUE9455561; US Dept. of Education FIPSE grants P116B970186 and P116B0000300; and US Dept. of Education grant R125K010106.